\documentclass[twocolumn,amsmath,amssymb,floatfix,prb,showpacs,footinbib,superscriptaddress]{revtex4}
\usepackage{amsmath,amssymb,natbib,bm,graphicx,url,psfrag,epsfig}
\usepackage[ansinew]{inputenc}
 \usepackage[normalem]{ulem} 
\newcommand{\be}{\begin{equation}}
\newcommand{\ee}{\end{equation}}
\newcommand{\bes}{\begin{equation}\begin{split}}
\newcommand{\ees}{\end{split}\end{equation}}

\newcommand{\ket}[1]{\left|\, #1 \, \right\rangle}

\begin{document}
\title{Vibrational absorption sidebands in the Coulomb blockade regime of single-molecule transistors}

\author{Matthias C.\ \surname{L\"uffe}}
\affiliation{Institut f\"ur Theoretische Physik, Freie
Universit\"at Berlin, Arnimallee 14, 14195 Berlin, Germany}
\author{Jens \surname{Koch}}
\affiliation{Department of Applied Physics, Yale University, New
Haven, CT 06520, USA}
\author{Felix \surname{von Oppen}}
\affiliation{Institut f\"ur Theoretische Physik, Freie
Universit\"at Berlin, Arnimallee 14, 14195 Berlin, Germany}

\date{September 27, 2007}
\begin{abstract}
Current-driven vibrational non-equilibrium induces vibrational sidebands in single-molecule transistors which arise from tunneling processes accompanied by {\em absorption} of vibrational quanta. Unlike conventional sidebands, these absorption sidebands occur in a regime where the current is nominally Coulomb blockaded. 
Here, we develop a detailed and analytical theory of absorption sidebands, including current-voltage characteristics as well as shot noise. We discuss the relation of our predictions to recent experiments.

\end{abstract}
\maketitle
\section{Introduction}

Single-molecule junctions and transistors have opened new vistas for quantum transport phenomena in nanostructures, which are associated with the coupling of the electronic degrees of freedom to collective modes of the molecule such as a molecular spin\cite{Zant,Ralph} or molecular vibrations.\cite{Nitzan} One of the most fascinating possibilities afforded by this coupling is that the current flow can drive the collective modes far out of thermal equilibrium, which will in turn act back on the current. This makes single-molecule transistors an interesting testbed for out-of-equilibrium quantum transport. Indeed, several recent experiments\cite{dekkernature,dekkerprb,Ruitenbeek} on single-molecule transistors have provided convincing evidence for out-of-equilibrium vibrations and theory has made numerous predictions\cite{Mitra04,Gorelik98,Koch05} for new effects arising from vibrational nonequilibrium. 

\begin{figure}[t]
\includegraphics[width=.95\columnwidth,clip=true]{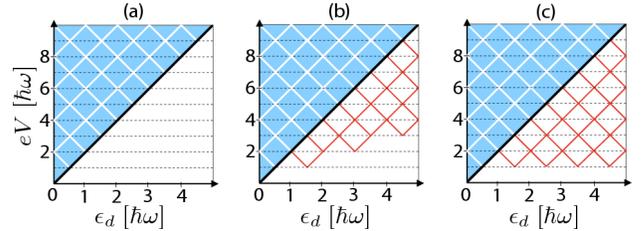}
\caption{(Color online) Schematic vibrational sidebands in $dI/dV$ vs.\ bias and gate-voltage plane for (a) fast, (b) intermediate, and (c) slow relaxation. Conventional emission sidebands in the sequential tunneling region (colored/shaded) are shown in white, absorption sidebands in red/gray. Horizontal dashed lines appear due to inelastic cotunneling, thick black lines due to sequential tunneling without change of the vibrational state.} 
\label{schematic}
\end{figure}

In the Coulomb blockade regime of transport through a single-molecule transistor, the coupling of the tunneling electrons to molecular vibrations induces vibrational sidebands.\cite{park,Mitra04} These sidebands occur for bias and gate voltages, where the Coulomb blockade is lifted, and arise from sequential tunneling processes which are accompanied by excitations of the molecular vibrations, cf.\ Fig.\ \ref{schematic}(a). If we consider tunneling processes which make the molecule switch between its neutral and singly-charged state, vibrational sidebands occur whenever $E_F+eV/2= \epsilon_d + n\hbar \omega$ due to tunneling into the molecule and $\epsilon_d = E_F-eV/2 +n \hbar \omega$ due to tunneling off the molecule. Here, $E_F$ is the Fermi energy of the leads in the absence of an applied bias $V$, $\epsilon_d$ denotes the energy of the relevant molecular orbital, and $\omega$ is the frequency of the vibrational mode. The positive integer $n$ labels the different vibrational {\it emission} sidebands, which typically weaken with increasing $n$.

These sidebands occur due to electron tunneling processes which are accompanied by {\em emission}, i.e.\ excitation, of vibrational quanta.
In the presence of current-driven out-of-equilibrium vibrations, it is natural to ask whether there exist additional features in transport through single-molecule transistors which originate from the {\em absorption} of vibrational quanta (similar to anti-Stokes lines in Raman spectroscopy). Indeed, it was recently observed \cite{koch7} in numerical simulations of transport through single-molecule junctions that vibrational non-equilibrium can induce additional absorption sidebands within the region where the current is nominally Coulomb blockaded. A schematic illustration of these absorption sidebands is shown in Fig.\ \ref{schematic}(b) and (c). 

It is the purpose of this work to present an analytical theory of these novel vibrational absorption sidebands. In Sec.\ \ref{sec:basicmechs}, we discuss the basic mechanism which leads to the formation of absorption sidebands and identify several qualitatively different regimes. Our analytical theory, including both coductance and shot noise, is presented in Sec.\ \ref{sec:analytic}. In Sec.\ \ref{sec:conclusions}, we discuss the relation of our results to recent experiments \cite{dekkernature,dekkerprb} and argue that observation of the vibrational absorption sidebands allows one to measure the vibrational relaxation time in single-molecule transistors. Some technical details have been relegated to an Appendix.

\section{Basic mechanisms\label{sec:basicmechs}}

\begin{figure}
    \centering
 \includegraphics[width=.85\columnwidth]{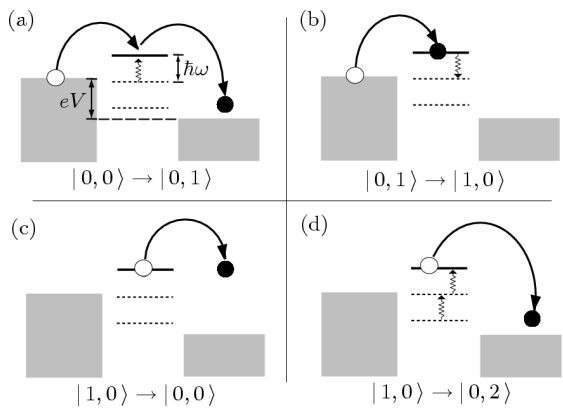}
    \caption{Relevant transport processes in the vicinity of the first absorption sideband for $\hbar\omega < eV < 2\hbar\omega$. In addition to elastic cotunneling (not shown), the following processes contribute to current flow: (a) Inelastic cotunneling with emission of one vibron; (b) population of the molecule by inelastic sequential tunneling with absorption of one vibron; depopulation by (c) elastic sequential tunneling or (d) inelastic sequential tunneling with emission of at most two vibrons.  Molecular states are symbolized by $|n,q\rangle$ where $n$ refers to the charge state and $q$ to the vibrational state.
\label{levelschema}}
\end{figure}

The absorption sidebands occur for gate and bias voltages where the current is nominally Coulomb blockaded. In other words, the energy level of the molecular orbital lies outside the bias window. This is illustrated in Fig.\ \ref{levelschema} where we depict the processes relevant in the vicinity of the first absorption sideband for $\hbar\omega < eV < 2\hbar\omega$.

The basic process which allows current to flow through the molecule is cotunneling in which an electron tunnels between the leads with intermediate virtual occupation of the molecule. Cotunneling necessarily leaves the vibrational state unchanged for bias voltages $eV<\hbar\omega$ (elastic cotunneling). Assuming temperatures $T\ll \hbar\omega$, vibrons can be excited once $eV>\hbar\omega$ [inelastic cotunneling, cf.\ Fig.\ \ref{levelschema} (a)]. In plots of the differential conductance of the single-molecule transistor, inelastic cotunneling due to excitation of vibrons leads to features at $eV=n\hbar\omega$ with $n=1,2\ldots$, independent of the gate voltage,\cite{inelastic,Mitra04} cf.\ Fig.\ \ref{schematic}.

If the vibrons excited by inelastic cotunneling decay only slowly, their presence opens a new transport channel which is responsible for the occurrence of vibrational absorption sidebands. Indeed, absorption of these vibrons by an electron may energetically allow {\it sequential} tunneling processes onto or off the molecule even though 
they are forbidden in the absence of vibron absorption [cf.\ Fig.\ \ref{levelschema} (b)]. For parameters in the vicinity of the coexistence of the neutral and the singly-charged molecule, the $n$th vibrational absorption sideband occurs when sequential tunneling becomes possible with absorption of $n\hbar\omega$. Setting $E_F=0$, the locations of the absorption sidebands are ${eV}/{2} + n\hbar \omega = \epsilon_d$ for tunneling onto the molecule (with $\epsilon_d>eV/2$) and $\epsilon_d + n\hbar \omega = - {eV}/{2}$ for tunneling off the molecule (with $\epsilon_d<-eV/2$). In addition, the condition $eV>\hbar \omega$ must be satisfied because vibrons can be absorbed only after they are first excited in inelastic cotunneling events. 

Rates for allowed sequential tunneling processes are lower order in perturbation theory in the molecule-lead hopping amplitude than electron cotunneling. Thus, these sequential tunneling rates $W_{\rm seq}$ are generally much larger than the cotunneling rates $W_{\rm cot}$. For this reason, we expect that (if energetically allowed), inelastic cotunneling events are rapidly followed by a sequential tunneling event involving vibron absorption. This scenario implies that three different regimes need to be distinguished: (i) fast vibrational relaxation: $1/\tau \gg W_{\rm seq} \gg W_{\rm cot}$; (ii) intermediate relaxation: $W_{\rm seq} \gg 1/\tau \gg W_{\rm cot}$; (iii) slow vibrational relaxation: $W_{\rm seq} \gg W_{\rm cot} \gg 1/\tau$. There are no absorption sidebands for fast vibrational relaxation. Here, we focus on the other two regimes of intermediate and slow vibrational relaxation, both of which exhibit such additional sidebands. Interestingly, there is a {\it qualitative} difference between (ii) and (iii), cf.\ Fig.\ \ref{schematic}(b) and (c): For intermediate relaxation, emitted vibrons will decay before the next inelastic cotunneling event, even in the absence of a sequential tunneling process. This implies that a sequentially tunneling electron can absorb $n\hbar\omega$ only if the preceding inelastic cotunneling event excited the molecular vibrations by $n\hbar\omega$. As a result, the $n$th absorption sideband occurs only for voltages $eV>n\hbar\omega$. In contrast, for slow vibrational relaxation, the vibronic state can be excited to levels beyond $eV$ by repeated inelastic cotunneling. Thus, higher absorption sidebands become visible as soon as inelastic cotunneling becomes possible for $eV>\hbar\omega$. 

This picture of transport in the regime of vibrational absorption sidebands has several additional consequences. 
(i) With increasing vibrational relaxation time, the current through the junction will increase since each inelastic cotunneling event will be followed by an increasing average number of sequentially tunneling electrons. (ii) The fact that inelastic cotunneling events are rapidly followed by sequentially tunneling electron(s) effectively implies electron bunching and thus super-Poissonian shot noise. (iii) This bunching becomes stronger for larger electron-vibron coupling $\lambda$. Therefore, vibrational absorption sidebands become more prominent with increasing $\lambda$. 
 
Vibrational absorption sidebands are amenable to an analytical description for both weak and strong electron-vibron interaction, $\lambda\ll 1$ and $\lambda\gg 1$.\cite{Strong} Results for these regimes give good qualitative insight also for the crossover region of intermediate coupling constants $\lambda\sim 1$, where an analytical description becomes more cumbersome. 

\section{Quantitative theory\label{sec:analytic}}

\subsection{Model and basic processes}

\label{basicprocesses}
In view of the generic nature of the phenomenon under discussion, our considerations are based on the Anderson-Holstein model\cite{glazman2,wingreen,Mitra04,flensb1} for a single-molecule device in which transport is dominated by one gate-tunable spin-degenerate molecular orbital $\epsilon_d$ coupled to a single harmonic vibrational mode with frequency $\omega$, as described by the Hamiltonian
\begin{align}
H_\text{mol}= \epsilon_d n_d + {U} n_{d\uparrow}n_{d\downarrow}+ \hbar \omega b^\dag b + \lambda \hbar \omega (b^\dag+b) n_d . \label{Hmol}
\end{align} 
Here, $U$ is the charging energy, the operator $d_\sigma$ annihilates an electron with spin projection $\sigma$ on the molecule, and $n_{d}= n_{d\uparrow}+n_{d\downarrow}=\sum_\sigma d_\sigma^\dag d_\sigma$ denotes the corresponding occupation-number operator. $b$ annihilates the vibrons. Tunneling between the molecule and  (Fermi-liquid) source and drain electrodes with Fermi energies $\mu_{L,R}=\pm eV/2$ is assumed to be weak with $\Gamma\ll T$. ($\Gamma$ is the tunnel broadening of the molecular orbital in the absence of the vibrational mode and $T$ the temperature.) 

Explicit results for the tunneling rates can be obtained by expanding the ${\cal T}$-matrix in the molecule-lead tunneling and applying Fermi's golden rule. To be specific, we focus on absorption sidebands in the Coulomb-blockade diamond where $n_d=0$. Cotunneling processes between leads $a$ and $b$ involving a change of the vibrational state from $q$ to $q^\prime$ occur with a rate (assuming $U\to\infty$)
\begin{align} 
W^{\rm cot}_{qq';ab}=~&\frac{s^{\rm cot}}{2 \pi \hbar} \Gamma_a \Gamma_b \int d\epsilon \left|\sum_{q"}\frac{M_{q' q''} M_{q q''}}{\epsilon - \epsilon_d + (q-q'')\hbar \omega}  \right|^2 \nonumber \\ & \times f_a(\epsilon) [1-f_b(\epsilon+[q-q']\hbar\omega)].
\end{align}
Here, $s^{\rm cot}= 2$ denotes a spin factor and $f_a(\epsilon)$ the Fermi distribution of lead $a$. 
Analytical results for the cotunneling rates are obtained using the integrals
\begin{eqnarray}
  &&\int d\epsilon \frac{f(\epsilon-E_1)[1-f(\epsilon-E_2)]}{(\epsilon-\epsilon_1)
  (\epsilon-\epsilon_2)} = \frac{n_B(E_2-E_1)}{\epsilon_1-\epsilon_2} \nonumber\\
  &&\,\,\,\,\,\,\,\,\times {\rm Re}\{\psi(E_{21}^+)-\psi(E_{22}^-)-\psi(E_{11}^+)+\psi(E_{12}^-)\} \\
  &&\int d\epsilon \frac{f(\epsilon-E_1)[1-f(\epsilon-E_2)]}{(\epsilon-\epsilon_1)^2} =
   \frac{n_B(E_2-E_1)}{2\pi T} \nonumber\\
  &&\,\,\,\,\,\,\,\,\times {\rm Im}\{\psi^\prime(E_{21}^+)-\psi^\prime(E_{11}^+)\} 
\end{eqnarray}
Here, the poles have been regularized in a standard fashion.\cite{averin,koch7} We used the notation $E_{ij}^\pm = \frac{1}{2}\pm \frac{i}{2\pi T} [E_i-\epsilon_j]$, the digamma function $\psi(z)$, and the Bose distribution $n_B(E)$.
The Franck-Condon matrix elements $M_{q_1 q_2}=\langle q_1|e^{\lambda(b-b^\dagger)}|q_2\rangle$ between vibrational states $|q\rangle$ have the analytical expression 
\begin{equation}
M_{q_1 q_2}=[{\rm sgn}(q_2-q_1)]^{q_1-q_2}
\lambda^{Q-q}e^{-\lambda^2/2}\left(\frac{q!}{Q!}\right)^{1/2}L^{Q-q}_{q}(\lambda^2)
\end{equation}
 in terms of generalized Laguerre polynomials $L^n_m(x)$ where $q={\rm min}(q_1,q_2)$, $Q={\rm max}(q_1,q_2)$. Vibration-assisted sequential tunneling onto the molecule involving lead $a$ and changing the vibrational state from $q$ to $q^\prime$ has the rate 
\begin{equation}
W^{01}_{q q';\,a}= s^{0\rightarrow 1}\Gamma_a  |M_{q q'}|^2 f_a (\epsilon_d+[q^\prime-q]\hbar\omega)
\end{equation}
 with the spin factors $s^{1\rightarrow 0}= 1$ and $s^{0\rightarrow 1}= 2 $. An analogous expression holds for $W^{1 0}_{q q';\,a}$. 
Below, we will drop the index ``$ab$'' on the cotunneling rates from source to drain (LR). We also abbreviate the total cotuneling rate as $W^\textnormal{cot}_{qq';{\rm tot}}=\sum_{a,b} W^\textnormal{cot}_{qq';ab}$. Sequential tunneling rates without lead index refer to the sum of the rates across both junctions, $W_{qq^\prime}^{10}=W_{q q^\prime;L}^{10}+W_{qq^\prime;R}^{10}$.

\subsection{Intermediate vibrational relaxation}

\subsubsection{Current}

For weak electron-vibron coupling $\lambda\ll 1$, crucial simplifications occur due to the Franck-Condon matrix elements favoring processes with a minimal change in the vibron occupation. This allows us to restrict attention to the occupation probabilities $P_q^n$ of a limited number of states $|n,q\rangle$ of the molecule. Here, $n$ denotes the charge state and $q$ the vibrational state. For example, consider the system with $\lambda\ll 1$ after an electron has sequentially tunneled into the molecular state $|1,0\rangle$ by vibron absorption. Tunneling out to the drain electrode can now be accompanied by emission of one or several vibrons [cf.\ Fig.\ \ref{levelschema}(d)]. However, to lowest order in $\lambda$, we only need to include the process with the final state $\ket{0,0}$ in which no vibron is excited [cf.\ Fig.\ \ref{levelschema}(c)]. 

We first focus on the $n$th vibrational absorption sideband, for $eV>n\hbar\omega$ and intermediate relaxation. The states $|0,0\rangle$, $|0,j  \le n\rangle$, and $|1,0\rangle$ need to be included to leading order. Solving the corresponding rate equations, we obtain for the stationary current (see appendix \ref{appendixCurrents} for details)
\begin{eqnarray}
  I = \sum_{j=0}^{n}eW_{0j}^{\rm cot} +  eW_{0n}^{\rm cot}
   \frac{W_{n0;L}^{01}}{W_{n0;L}^{01}+\frac{1}{\tau}} \frac{W_{00;R}^{10}}{W_{00}^{10}}. 
  \label{Isideband}
\end{eqnarray}
The first term on the right-hand side accounts for contributions of elastic and inelastic cotunneling. The second term is responsible for the $n$th absorption sideband. Here, the inelastic cotunneling rate $W_{0n}^{\rm cot}$ reflects that the corresponding process opens the absorption-assisted sequential tunneling channel by which additional electrons can pass through the molecule. The first fraction arises from the competition between sequential tunneling and vibrational relaxation. It is the rate $W_{n0,L}^{0 1}$ which turns on at the $n$th absorption sideband and thus leads to a peak in the differential conductance $dI/dV$ vs.\ bias and gate voltage when ${eV}/{2} + n\hbar \omega = \epsilon_d$. The second fraction describes that an additional electron passes through the molecule only when the electron on the molecule sequentially tunnels out to the right lead (drain).

For strong electron-vibron coupling $\lambda\gg 1$, the Franck-Condon matrices favor transitions where the final state has the largest accessible vibron occupation.\footnote{Strictly speaking, this statement holds true in the relevant range of transistions between low-lying vibrational states below the Franck-Condon parabola.} Unlike for weak coupling, this allows for longer sequences of more than two sequential tunneling events. This is because tunneling out from the molecule will now preferredly lead to final states with vibronic excitations [e.g., process (d) in Fig.\ \ref{levelschema} will be dominant with respect to process (c)]. Hence, the restriction to only a small set of states is no longer feasible. Instead, the system can acquire a large number of vibrons. The sequence of sequential tunneling events is ultimately cut off by subdominant sequential processes or vibrational relaxation. In appendix \ref{appendixCurrents}, this is illustrated in detail for the first vibrational absorption sideband $\hbar \omega < e V < 2 \hbar \omega$ where we find
\begin{align}
I&~=~eP_0^0 (W_{00;LR}^\mathrm{cot}+W_{01;LR}^\mathrm{cot})+e\sum_{i=0}^{\infty} P_i^1 \;W_{i, i+2;R}^{10} \nonumber\\&~=~I_\mathrm{el}+eW_{01;LR}^\mathrm{cot} +e W_{01;LR}^\mathrm{cot}\,\frac{ W_{10;L}^{01}}{W_{10;L}^{01}+\frac{1}{\tau} } \nonumber \\
& \times\left[1 + \frac{1}{W_{02;R}^{10}}\sum_{i=1}^{\infty}\,\prod_{k=1}^{i}\, \frac{W_{k+1,k;L}^{01}}{W_{k,k+2;R}^{10}}\; \frac{W^{10}_{k-1,k+1;R}W_{i, i+2;R}^{10}}{W^{10}_{k+1,k;L}+\frac{1}{\tau}}\right].
  \label{Isidestrong}
\end{align}
Here, the first term describes the elastic cotunneling and the second term the inelastic cotunneling which initiates the series of sequential events described by the third term. The expression Eq.\ \eqref{Isidestrong} makes it again explicit that the first vibrational sideband emerges for intermediate relaxation, i.e., once the {\it sequential} tunneling rate $W_{10;L}^{01}$ becomes faster than the vibrational relaxation $1/\tau$.

\subsubsection{Fano factor}

An interesting feature of Eqs.\ (\ref{Isideband}) and (\ref{Isidestrong}) is that the current is partially vibron assisted in the sense that it increases with increasing vibrational relaxation time $\tau$. This is because the vibron-assisted sequential tunneling processes which may follow inelastic cotunneling event, transfer additional electrons. Clearly, the number of additional electrons passing through the molecule increases when
going from fast to intermediate vibrational relaxation. E.g., for $\lambda\ll 1$ the average number of electrons passing per inelastic cotunneling event is 1 for fast relaxation, but $3/2$ (symmetric junction) or $2$ (strongly asymmetric STM configuration) for intermediate relaxation. 

Since sequential tunneling proceeds on a much faster timescale, these considerations imply electron bunching and super-Poissonian shot noise. For $\lambda\ll 1$, the corresponding Fano factor $F$ can be obtained from $F=(I_1+2I_2)/(I_1+I_2)$, where $I_1$ ($I_2$) denotes the contribution to the current due to single-electron (two-electron) processes. For intermediate vibrational relaxation near the $n$th absorption sideband with $eV>n\hbar\omega$, we obtain  
\begin{equation}
  F = 1 + 2\frac{W_{0 n}^{\rm cot}}{W_{00}^{\rm cot}}\, \frac{W_{00;R}^{1 0}}{W_{00;L}^{1 0}+
   W_{00;R}^{1 0}}.
\label{Fano}
\end{equation}
Due to the first fraction, the Fano factor for weak coupling will take a value only slightly above one. 

\begin{figure}[t]
\includegraphics[width=1\columnwidth]{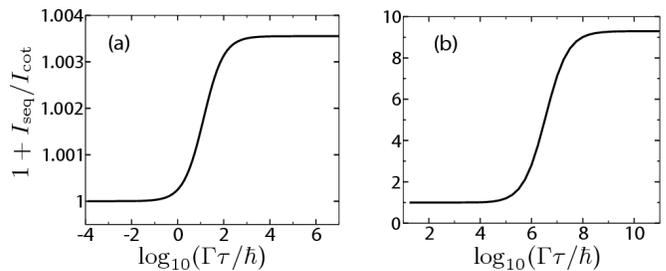}
\caption{The quantity $\tilde F=1+I_{\rm{seq}}/I_{\rm{cot}}$ as defined in the text vs.\ the relaxation time for (a) weak ($\lambda=0.2$) and (b) strong ($\lambda=4$) electron-vibron coupling. ($V=1.7 \,\hbar\omega;\, \epsilon_d=1.6 \,\hbar\omega$)
\label{quasiff}} 
\end{figure}

Much larger Fano factors are to be expected for strong electron-vibron interaction where cotunneling events can be followed by a long sequence of vibron assisted sequential tunneling events. An approximate measure of the Fano factor is given by $\tilde F=1+I_{\rm seq}/I_{\rm cot}$, where $I_{\rm seq}$ ($I_{\rm cot}$) denotes contribution to the current due to sequential (cotunneling) processes. This is an instructive measure for the average number of sequential tunneling events following a cotunneling process. We plot numerical results for $\tilde F$ as a function of $\log_{10}(\Gamma\tau/\hbar)$ for both weak coupling [Fig.~\ref{quasiff}(a)] and strong coupling [Fig.~\ref{quasiff}(b)]. In both cases, $\tilde F$ features a step-like increase at the crossover between fast and intermediate relaxation but the step height is much greater for strong electron-vibron coupling. (Note the different scales on the vertical axes in Fig.~\ref{quasiff}.)

\subsection{Slow vibrational relaxation}

For voltages $eV< n\hbar\omega$, the $n$th vibrational absorption sideband becomes visible only for slow vibrational relaxation, since the molecular vibrations must be excited to the $n$th level by \emph{repeated} inelastic cotunneling processes. To make this point quantitative, we consider the second absorption sideband for voltages $\hbar\omega<eV<2\hbar\omega$ and weak electron-vibron coupling $\lambda\ll 1$. (Analytical results for strong coupling become more cumbersome and will not be discussed here.) We find (see appendix \ref{appendixCurrents} for details)
\begin{eqnarray}
    I = I_{\rm cot}+ eP_0^0 
   W_{0 1}^{\rm cot}\frac{W_{12}^{\cot}}{D}\left[2+\frac{W_{00;R}^{1 0}}{W_{00}^{1 0}} \right],
\label{Isideband2}
\end{eqnarray}
where $I_{\rm cot}= e P_0^0 W_{0 0}^{\rm cot}+eW_{0 1}^{\rm cot} P_0^0 [1+{W_{10}^{\rm cot}}/{D}+{W_{11}^{\rm cot}}/{D}]$ collects cotunneling processes without subsequent sequential tunneling, $D=W_{12}^{\rm cot}+W_{10;\mathrm{tot}}^{\rm cot}+1/\tau$, and $P_0^0=D/[D+W_{0 1}^{\rm cot}]$. 

The product of two cotunneling rates appearing in the second term of Eq.\ (\ref{Isideband2}) reflects the need for two subsequent inelastic cotunneling events to excite the vibrational level $q=2$. Due to the denominator $D$ the second absorption sideband appears for these voltages once vibrational relaxation becomes slow compared to inelastic cotunneling.

\section{Conclusions\label{sec:conclusions}}

Vibrational absorption sidebands exhibit a rich phenomenology due to the presence of three relevant time scales derived from the rates for cotunneling, sequential tunneling, and vibrational relaxation. This leads to the existence of different regimes, and as a consequence
the observation of vibrational absorption sidebands can be employed to measure the vibrational relaxation time. This is particularly promising in scanning tunneling microscope measurements where the current level can be tuned by changing the tip-to-molecule distance. In this case, the vibrational absorption sidebands should emerge only once the current exceeds a certain threshold current, and one could extract the vibrational relaxation time $\tau$ in two different ways:   

(i) While vibrational absorption sidebands emerge already when $1/\tau \approx W_{\rm seq}$, one suggestion is to focus on higher-order ($n>1$) absorption sidebands for $\hbar\omega < eV <2 \hbar\omega$. These sidebands appear only once the vibrational relaxation rate becomes comparable with inelastic cotunneling rates. Assuming electron-vibron coupling constants of order unity, the current is given by the inelastic cotunneling rate (multiplied by $e$) up to a numerical factor of order unity. If these sidebands appear at a threshold current of order $I_c$, the vibrational relaxation rate can be estimated from $1/\tau \sim I_c/e$. For electron-vibron coupling constants $\lambda$ which are not too close to unity, a more accurate quantitative comparison with the theory developed here should provide a good estimate for $1/\tau$. 

(ii) While suggestion (i) requires slow vibrational relaxation, an alternative approach which applies also to intermediate vibrational relaxation consists of a separate measurement of the sequential tunneling rate $W_{\rm seq}$. This is readily done by measuring the current at the same bias, but at a gate voltage such that the single-molecule junction is outside the Coulomb-blockaded regime, where the current is dominated by sequential tunneling. 

Indeed, recent experiments \cite{dekkernature,dekkerprb} have extracted vibrational relaxation times for suspended single-wall carbon nanotubes in a related way and found them to be of the order of 10ns. However, it is important to understand that the regime discussed in the present work differs from that investigated in these experiments in an essential way. The vibrational absorption sidebands discussed here occur for gate and bias voltages where the current is nominally Coulomb blockaded. By contrast, the experiments of Refs.\ \onlinecite{dekkernature} and \onlinecite{dekkerprb} are performed at high bias, well outside the Coulomb blockade region, where several ``molecular orbitals" are located within the bias window. 

Finally, we remark that the magnitude of the vibrational absorption sidebands may also allow access to the ``effective 
temperature" of the out-of-equilibrium molecular vibrations which is an important characteristic of the single-molecule junction but has so far remained outside experimental reach.

\begin{acknowledgments}
We acknowledge useful discussions with N.\ Pascual. This work was supported in part by Sfb 658, SPP 1243 and Yale University (JK). 
\end{acknowledgments}

\begin{appendix}

\section{Calculation of the currents from reduced rate equations}
\label{appendixCurrents}
Here, we sketch the derivations for the current expressions Eqs.\ \eqref{Isideband}, \eqref{Isidestrong}, and \eqref{Isideband2}, valid in different regimes of relaxation strength and electron-vibron coupling. The general procedure is as follows: For each particular regime, we identify the relevant molecular states and transitions between them. The corresponding rates are calculated from a ${\cal T}$-matrix expansion as discussed in section \ref{basicprocesses}. Many rates are found to be essentially zero, either because the process is not allowed at small temperatures ($T \ll eV,\hbar\omega$), or because the vibronic transition is suppressed due to the Franck-Condon matrix element. Eliminating these processes reduces the system of rate equations determining the stationary occupation probabilities $P^n_q$ according to 
\begin{align}\label{rateeq}
0~=~\frac{dP^n_q}{dt}~=~&\sum_{n', q'}  \left[ P^{n'}_{q'} W^{n'  n}_{q'  q} - P^{n}_{q} W^{n n'}_{q q'} \right]\nonumber\\
&-\frac{1}{\tau}[{\textstyle P^n_q- P^\text{eq}_q \sum_{q'} P^n_{q'}}].
\end{align} 
For simplicity, we have used the notation $W^{n n}_{q q'}$ for cotunneling rates. The last term in Eq.~\eqref{rateeq} models relaxation towards the equilibrium vibron distribution $P^\text{eq}_q$ with time constant $\tau$.
The stationary current is then given by the sum of the contributions from sequential tunneling and cotunneling,
\begin{align}\label{current}
I~=&~e\sum_{n,q,q'}P^{n}_{q} \left[W^{n (n-1)}_{q q';R} - W^{n (n+1)}_{q q';R} \right]\nonumber\\&+e\sum_{n,q,q'}P^{n}_{q} \left[W^{n n}_{q q';LR} - W^{n n}_{q q';RL} \right].
\end{align}

\subsection{Intermediate relaxation, weak electron-vibron coupling}
\begin{figure}[t]
\includegraphics[width=1.0\columnwidth,clip=true]{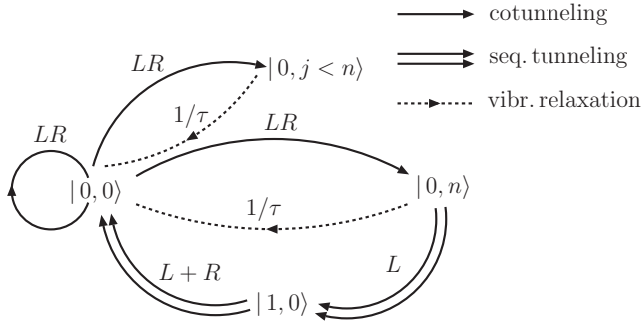}
\caption{Relevant states and processes for intermediate relaxation and weak electron-vibron coupling in the vicinity of the $n$th absorption sideband. Single solid lines represent cotunneling transitions, double lines sequential tunneling and dashed lines vibrational relaxation.} 
\label{sketchWeakInterm}
\end{figure}

We first consider the $n$th absorption sideband for $\lambda \ll 1$ and intermediate relaxation. The states and processes relevant in this regime are depicted in Fig.~\ref{sketchWeakInterm}. 
The left and upper right branches of the scheme (originating from the ground state) represent elastic and inelastic cotunneling channels. These compete with the series of processes (lower right branch) which is responsible for the formation of the $n$th sideband. 
The reduced system of rate equations which describes this interesting part  of the scheme is 
 \begin{align}
\begin{pmatrix}
-W_{0n;LR}^{\textnormal{cot}} &1/\tau & W_{00}^{10} \vspace{0.2 cm}  \\ 
W_{0n;LR}^{\textnormal{cot}} & -W_{n0;L}^{01}-1/\tau & 0 \vspace{0.2 cm}\\ 
 0&W_{n0;L}^{01} &-W_{00}^{10} 
\end{pmatrix}
\begin{pmatrix}
 P_0^0 \vspace{0.2 cm}\\
 P_n^0 \vspace{0.2 cm}\\
 P_0^1
\end{pmatrix}
=\begin{pmatrix}
 0 \vspace{0.2 cm}\\
 0 \vspace{0.2 cm}\\
 0
\end{pmatrix}.\label{lgs}
\end{align}
Here we have used Eq.~\eqref{rateeq} with $P^\text{eq}_0 \approx 1$, valid for $T \ll \hbar \omega$.
>From this we obtain the probabilities
\begin{align}
P_n^0&=\frac{W_{0n;LR}^{\textnormal{cot}}}{W_{n0;L}^{01} +1/\tau} P_0^0,\label{secondrow}\\
P_0^1&=\frac{W_{n0;L}^{01}}{W_{00}^{10}} \frac{W_{0n;LR}^{\textnormal{cot}}}{W_{n0;L}^{01} +1/\tau} P_0^0.\label{thirdrow}
\end{align}
Insertion of Eq.~\eqref{thirdrow} into the equation for the current, 
\begin{align}
I&=P_0^0\sum_{j=0}^{n-1}e W_{0jLR}^{\textnormal{cot}}+ P_0^0 e W_{0n;LR}^\textnormal{cot}+P_0^1e W_{00;R}^{10}
\end{align}
with $P_0^0 \approx 1- W_\textnormal{cot}/(W_\textnormal{seq}+ 1/\tau) \approx 1$ yields Eq.~\eqref{Isideband}. Intuitively, the last approximation can also be understood from the fact that the system will predominantly occupy the ground state. While inelastic cotunneling makes the system leave the ground state, such excursions only happen on a very short time scale, limited by the quick return to the ground state via sequential tunneling or vibrational relaxation.

\subsection{Intermediate relaxation, strong electron-vibron coupling}
\begin{figure}[t]
\includegraphics[width=.9\columnwidth,clip=true]{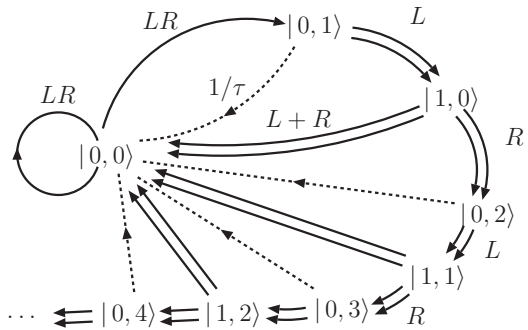}
\caption{Dynamics leading to the first absorption sideband (in the bias range $\hbar\omega < eV < 2\hbar\omega$) for intermediate relaxation and strong electron-vibron coupling.} 
\label{sketchStrongInterm}
\end{figure}

We now turn to the limit $\lambda \gg 1$ in the intermediate relaxation regime. We focus on the first absorption sideband in the bias range $\hbar\omega < eV < 2\hbar\omega$. The corresponding dynamics is sketched in Fig.~\ref{sketchStrongInterm}.
Unlike in the $\lambda \ll 1$ case, here, sequential tunneling out of the molecular orbital predominantly leads to  highly excited vibronic final states. From there, the molecule can be repopulated by inelastic sequential tunneling. As a result, the initiating inelastic cotunneling event may be followed by a chain of several sequential processes. The reduced system, depicted in Fig.~\ref{sketchStrongInterm}, includes not only processes of lowest order in $1/\lambda$ 
but also higher order transitions $\ket{1,q} \rightarrow \ket{0,0}$ which serve as limiting processes for the chain of sequential transitions. 
By contrast, transitions $\ket{1,q} \rightarrow \ket{0,q'}$ with $0<q'<q+2$ are neglected since they produce only a higher-order correction to the sequential current.
The stationary current is
\begin{align}
I&=P_0^0 e (W_{00;LR}^\mathrm{cot}+W_{01;LR}^\mathrm{cot})+\sum_{i=0}^{\infty} P_i^1 e (W_{i, i+2;R}^{10}+W_{i0;R}^{10}).\label{currentstrong}
\end{align}
Here, the probabilities $P_i^1$ are determined from the 
rate equations corresponding to Fig.~\ref{sketchStrongInterm}. The resulting system of linear equations in this case is infinite. However, its simple structure allows for a recursive solution of $P_i^1$ in terms of $P_{i+1}^0$, which can in turn be expressed in terms of $P_{i-1}^1$ and so on. For strong electron-vibron coupling, we have
$W_{i, i+2;R}^{10}\gg W_{i0;R}^{10} $, so that the second term in the sum in Eq.~\eqref{currentstrong} may be neglected. Approximating $P_0^0\approx1$, we arrive at Eq.~\eqref{Isidestrong}.

\subsection{Slow relaxation, weak electron-vibron coupling}
\begin{figure}[t]
\includegraphics[width=.8\columnwidth,clip=true]{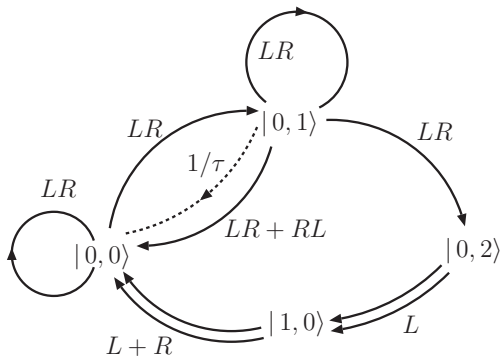}
\caption{Relevant states and transitions at the second absorption sideband (in the bias range $\hbar\omega < eV < 2\hbar \omega$) for slow relaxation and weak electron-vibron coupling.} 
\label{sketchWeakSlow}
\end{figure}

We finally turn to the regime of slow relaxation and weak electron-vibron coupling. We focus on the second absorption sideband in the bias range $\hbar\omega < eV < 2\hbar \omega$. As shown in Fig.~\ref{sketchWeakSlow}, at least two successive cotunneling transitions are required to open the sequential channel which generates the absorption sideband. 
The sketch is easily translated into a system of rate equations. Solving for the occupation probabilities and using $P_0^0+P_1^0\approx 1$ (up to corrections of order $W_\textnormal{cot}/W_\textnormal{seq}$) leads to Eq.~\eqref{Isideband2}.

\end{appendix}

\vspace{0.2 cm}

\begin{thebibliography}{xxxxxxx}
\bibitem{Strong} For $\lambda\gg1$, it may happen that $W_{\rm seq}\gg W_{\rm cot}$ is no longer valid, see Ref.\  \onlinecite{koch7}. We do not consider this situation.
\end{thebibliography}
\end{document}